\documentclass[10pt,twoside,a4paper]{amsart}
\usepackage{amsmath}
\usepackage{amsfonts}
\usepackage{amssymb}
\usepackage{amsthm}
\usepackage{newlfont}
\usepackage{graphicx}
\usepackage{amscd}

\theoremstyle{plain}
\newtheorem{Th}{Theorem}
\newtheorem{Cor}[Th]{Corollary}
\newtheorem{Lem}[Th]{Lemma}
\newtheorem{Prop}[Th]{Proposition}
\theoremstyle{definition}
\newtheorem{Def}{Definition}
\theoremstyle{remark}
\newtheorem*{Rem}{Remark}
\numberwithin{equation}{section}
\newcommand{\PP}{{\mathbb P}}
\newcommand{\KK}{{\mathbb K}}
\newcommand{\RR}{{\mathbb R}}
\newcommand{\EE}{{\mathbb E}}
\newcommand{\CC}{{\mathbb C}}
\newcommand{\ZZ}{{\mathbb Z}}

\newcommand{\cQ}{{\mathcal Q}}

\newcommand{\qcr}{\mathrm{cr}}

\newcommand{\bx}{{\boldsymbol x}}

\newcommand{\by}{{\boldsymbol y}}

\begin{document}

\title
{Generalized Isothermic Lattices}

\author[Adam Doliwa]{Adam Doliwa$^\ddagger$}
\thanks{$\ddagger$ Supported in part by the DFG Research Center MATHEON}

\address{Wydzia{\l} Matematyki i Informatyki,
Uniwersytet Warmi\'{n}sko-Mazurski w Olsztynie,
ul. \.{Z}o{\l}nierska 14, 10-561 Olsztyn, Poland}


\email{doliwa@matman.uwm.edu.pl}


\begin{abstract}

\noindent We study multidimensional
quadrilateral lattices satisfying simultaneously two 
integrable constraints: a quadratic constraint and the projective Moutard
constraint. 
When the lattice is two dimensional and the quadric under consideration is the
M\"{o}bius sphere one obtains, after the stereographic projection, 
the discrete isothermic surfaces defined by Bobenko
and Pinkall by an algebraic constraint imposed on the (complex) cross-ratio of
the circular lattice. We derive the analogous condition for our generalized
isthermic lattices using Steiner's projective structure of conics and we present
basic geometric constructions which encode integrability of the lattice. In
particular we introduce the Darboux transformation of the generalized isothermic
lattice and we derive
the corresponding Bianchi permutability principle.
Finally, we study two dimensional generalized isothermic lattices, in particular
geometry of their initial boundary value problem.
\\


\end{abstract}
\maketitle

\section{Introduction}
\subsection{Isothermic surfaces}
In the year 1837 Gabriel Lam\'{e} presented \cite{Lame-isoth}
results of his studies on distribution
of temperature in a homogeneous solid body in thermal equilibrium. 
He was interested, in particular, in
description of the \emph{isothermic surfaces}, i.e. surfaces of constant 
temperature within the body; 
notice that his definition makes sense only for families of surfaces, 
and not for a single surface. 
Then he found a condition 
under which one parameter family of 
surfaces in (a subset of) $\EE^3$ consists of isothermic surfaces,
and showed (for details see \cite{Lame} or~\cite{DarbouxOS}) 
that the three families of confocal quadrics, which provide elliptic
coordinates in $\EE^3$, meet that criterion. 
Subsequently, he proposed to
determine all triply orthogonal systems composed by three isothermic families
(triply isothermic systems). 
Such a program was fulfiled by Gaston Darboux~\cite{DarbouxOS} (see also
\cite{Eisenhart-SV}). 

Another path of research was initiated by Joseph Bertrand~\cite{Bertrand} who
showed that the surfaces of triply isothermic systems are divided by their lines 
of curvature into "infinitesimal squares", or in exact terms, they allow for
conformal curvature parametrization. This definition of isothermic
surfaces (or surfaces of isothermic curvature lines), which can be 
applied to a single surface, was commonly accepted
in the second half of the XIX-th century
(see~\cite{Bianchi,DarbouxIV}). We mention that the minimal surfaces 
and the constant mean curvature surfaces are particular examples of 
the isothermic surfaces.
The theory of 
isothermic surfaces was one of the most favorite subjects of study among 
prominent  geometers of that period. Such surfaces exhibit particular properties, for example
there exists a transformation, described by Gaston Darboux in~\cite{Darboux-transf}, which produces 
from a given isothermic surface a family of new surfaces of the same type.

The Gauss-Mainardi-Codazzi equations for isothermic surfaces constitute a
nonlinear system generalizing the $\mathrm{sinh}$-Gordon equation (the latter governs 
the constant mean curvature surfaces),
and the Darboux transformation can be interpreted as B\"{a}cklund-type
transformation of the system. Soon after that Luigi Bianchi
showed~\cite{Bianchi-isoth} that two Darboux transforms of a given isothermic surface 
determine \emph{in algebraic terms} new isothermic surface being their simultaneous
Darboux transform.
The Bianchi permutability principle can be
considered as a hallmark of integrability (in the sense of soliton theory) of
the above-mentioned system. Indeed, the isothermic surfaces were reinterpreted 
by Cie\'{s}li\'{n}ski, Goldstein and Sym \cite{CGS} within the 
theory of soliton surfaces~\cite{Sym}. 
More information on isothermic surfaces and their history the Reader can find
in the paper
of Klimczewski, Nieszporski and Sym \cite{KNS}, where also a more detailed description 
of the relation between the "ancient"
differential geometry and the soliton theory is given, and in books by Rogers and Schief
\cite{RogersSchief} and by Hertrich-Jeromin \cite{UHJ}.

\subsection{Discrete isothermic surfaces and discrete integrable geometry}
In the recent studies of the
relation between geometry and the integrable
systems theory a particular attention is payed to discrete 
(difference) integrable
equations and the corresponding discrete surfaces or lattice 
submanifolds. 
Also here the discrete analogs of
isothermic surfaces played a prominent role in the development of the subject.
Bobenko and Pinkall \cite{BP2} introduced the integrable discrete analogoue
of isothermic surfaces as mappings built of "conformal squares", i.e., maps
$\bx:\ZZ^2\to\EE^3$ with all
elementary quadrilaterals circular, and such that the complex cross-ratios 
(with the plane of a
quadrilateral identified with the complex plane $\CC$)
\[ q(m,n) = \qcr\big(\bx(m,n),\bx(m+1,n+1);\bx(m+1,n),\bx(m,n+1) \big)_\CC
\]
are equal to $-1$. 
Soon after that it turned out \cite{BP-V} that it is more convenient to allow
for the cross-rations to satisfy the constraint
\begin{equation} \label{eq:isoth-cr}
q(m,n) q(m+1,n+1) = q(m+1,n)q(m,n+1).
\end{equation}
Then the cross-ratio is a ratio of functions of single variables,
which corresponds to allowed re\-pa\-ra\-met\-ri\-za\-tion of the 
curvature coordinates 
on isothermic surfaces. 

After the pioneering work of Bobenko and Pinkall, which was an important 
step in building the geometric
approach to integrable discrete equations (see also 
\cite{DS-AL,BP1,DCN} and older results of the difference geometry
(\emph{Differenzengeometrie}) summarized in Robert Sauer's books
\cite{Sauer2,Sauer}), 
the discrete isothermic surfaces and their Darboux
transformations
were studied in a number of papers \cite{HeHP,Ciesl,Schief-C}. Distinguished
integrable reductions of isothermic lattices are the discrete constant mean
curvature surfaces or the discrete minimal surfaces \cite{BP-V,UHJ}. It should be
mentioned that the complex cross-ratio condition \eqref{eq:isoth-cr} was 
extended
to circular lattices of dimension three \cite{BP-V,Ciesl} 
placing the Darboux
transformations of the discrete isothermic surfaces on equal footing with the
lattice itself.  

\begin{figure}
\begin{center}
\includegraphics{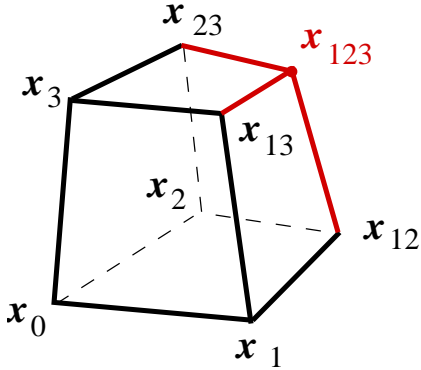}
\end{center}
\caption{The geometric integrability scheme}
\label{fig:TiTjTkx}
\end{figure}

In the present-day approach to the relation between
discrete integrable systems and 
geometry \cite{DS-EMP,BobSur} the key
role is played by the integrable discrete analogue of conjugate 
nets -- multidimensional lattices of planar quadrilaterals 
(the quadrilateral lattices) \cite{MQL}. These are
maps $x:\ZZ^N\to\PP^M$ of $N$-dimensional integer lattice 
in $M\geq N$-dimensional projective space with all
elementary quadrilaterals planar. Integrability of such lattices (for $N>2$)
is based on
the following elementary geometry fact (see Figure~\ref{fig:TiTjTkx}).
\begin{Lem}[The geometric integrability scheme] \label{lem:gen-hex}
Consider points $x_0$, $x_1$, $x_2$ and $x_3$ in general position in $\PP^M$,
$M\geq 3$. On
the plane $\langle x_0, x_i, x_j \rangle$, $1\leq i < j \leq 3$ choose a point
$x_{ij}$ not on the lines  $\langle x_0, x_i \rangle$, $\langle x_0,x_j
\rangle$ and $\langle x_i, x_j \rangle$. Then there exists the
unique point $x_{123}$
which belongs simultaneously to the three planes 
$\langle x_3, x_{13}, x_{23} \rangle$,
$\langle x_2, x_{12}, x_{23} \rangle$ and
$\langle x_1, x_{12}, x_{13} \rangle$.
\end{Lem}
Various constraints compatible with the
geometric integrability scheme define
integrable reductions of the quadrilateral lattice. It turns out that such
geometric notion of integrability very often associates 
integrable reductions of the quadrilateral lattice
with classical theorems of incidence geometry. 
We advocate this point of view in the present paper.

Among basic reductions of the 
quadrilateral lattice the so called quadratic
reductions \cite{q-red} play a distinguished role.
The lattice vertices are then contained in a 
hyperquadric (or in the intersection of several of
them). Such reductions of the 
quadrilateral lattice
can be often associated with various subgeometries of the projective geometry,
when the quadric plays the role of the absolute of the geometry
(see also corresponding remarks in \cite{q-red,W-cong}).
In particular, when the hyperquadric is the M\"{o}bius hypersphere 
one obtains, after the stereographic projection,
the circular lattices \cite{Bobenko-O,CDS,DMS,KoSchief2}, which are the 
integrable discrete analogue of submanifolds of
$\EE^M$ in curvature line parametrization. 
Because of the M\"{o}bius 
invariance of 
the complex cross-ratio it is also more convenient to consider (discrete) 
isothermic surfaces in the M\"{o}bius sphere (both dimensions of the 
lattice and of the sphere
can be enlarged) keeping the "cross-ratio definition". 

For a person trained in the projective geometry it is more or less 
natural to generalize the M\"{o}bius geometry approach to discrete isothermic
surfaces (lattices) in quadrics replacing the  M\"{o}bius sphere by a quadric, 
and correspondingly,
the complex cross-ratio by the Steiner cross-ratio 
of four points of a conic being intersection of the quadric by the plane of 
elementary quadrilateral of the quadrilateral lattice.  
However the "cross-ratio point of view" doesn't answer the crucial
question about integrability (understood as compatibility of the 
constraint with the geometric integrablity scheme) of such
discrete isothermic surfaces in quadrics. Our general methodological principle 
in the integrable
discrete geometry, applied successfuly earlier, for example in
\cite{DS-sym,DNS-Bianchi}, which we would like to follow here 
is (i) to isolate basic reductions of the quadrilateral 
lattice and then (ii) to 
incorporate other geometric systems into the theory considering them as 
superpositions of the basic
reductions. 

In this context we would like to recall
another  equivalent
characterization of the classical isothermic
surfaces which can be found in the classical monograph of 
Darboux~\cite[vol. 2, p. 267]{DarbouxIV}:
{\it
Les cinq coordonn\'{e}es pentasph\'{e}riques d'un point de toute surface
isothermique consid\'{e}r\'{e}es comme fonctions des param\`{e}tres $\rho$ et
$\rho_1$ des lignes de courbure satisfont \`{a} une \'{e}quation lin\'{e}aire du
second ordre dont les invariants sont \'{e}gaux.
Inversement, si une \'{e}quation de la forme
\begin{equation} \label{eq:Moutard-diff}
\frac{\partial^2\theta}{\partial \rho \partial \rho_1} = \lambda \theta
\end{equation}
ou, plus g\'{e}n\'{e}ralement, une \'{e}quation \`{a} invariants \'{e}gaux,
admet cinq solutions particuli\`{e}res $x_1$, $x_2$, \dots , $x_5$ li\'{e}es 
par
l'\'{e}quation
\begin{equation}
\sum_{1}^{5} x_i^2 = 0,
\end{equation}
les quantit\'{e}s $x_i$ sont les coordonn\'{e}es pantasph\'{e}riques qui
d\'{e}finissent une surface isothermique rapport\'{e}e \`{a} ses lignes de
courbure.
}

In literature there are known two (closely related) discrete integrable 
versions
(of Nimmo and Schief \cite{NiSchief} and of Nieszporski \cite{Nieszporski-dK}) 
of the Moutard equation \eqref{eq:Moutard-diff}. It turns out that for our
purposes it suits the discrete Moutard equation proposed in~\cite{NiSchief}. 
Its projectively invariant geometric characterization has been 
discovered \cite{DGNS} only recently (for geometric meaning
of the adjoint Moutard equation of Nieszporski in terms of the so called 
Koenigs lattice see \cite{Dol-Koe}). Indeed, it turns out that
the generalized
isothermic lattices can be obtained by adding to the quadratic
constraint the projective Moutard constraint. Finally, the quadratic reduction 
and the Moutard reduction, when applied simultaneously, give 
\emph{a posteriori} the cross-ratio condition \eqref{eq:isoth-cr}. 

In fact, the direct algebraic discrete counterpart of the above description of
the isothermic surfaces, i.e. existence of the light-cone lift which
satisfies the (discrete) Moutard equation, appeared first in a preprint by
Bobenko and Suris \cite{BobSur}.
However, the pure geometric characterization of the discrete isothermic surfaces
was not given there. 

Because integrability of the discrete Moutard equation can be seen 
better when one considers a system of such equations for multidimensional
lattices, there was a need to find the projective geometric
characterization of the system. The corresponding reduction of the quadrilateral
lattice was called in \cite{BQL}, because of its connection with the 
discrete BKP equation, the B-quadrilateral lattice (BQL).
In fact, research in this direction prevented me
from publication of the above mentioned generalization of discrete isothermic
surfaces, announced however in my talk during the 
Workshop "Geometry and Integrable Systems"
(Berlin, 3-7 June 2005). I suggested also there that the discrete
S-isothermic surfaces
of Tim Hoffmann \cite{Hoffmann} (see also \cite{BobSur})
should be considered as an example of the
generalized isothermic lattices where the quadric under consideration is the 
Lie quadric.
The final results of my research on generalized
isothermic lattices were presented on the Conference "Symmetries and
Integrability of Difference Equations VII" (Melbourne, 10-14 July 2006).

When my paper was almost ready there appeared the preprint of Bobenko and Suris
\cite{BobSur-gen-isoth} where similar ideas were presented in application to the
sphere (M\"{o}bius, Laguerre and Lie) geometries. I would like also to point
out a recent paper by Wallner and Pottman \cite{WP} devoted, among others, to
discrete isothermic surfaces in the Laguerre geometry.

\subsection{Plan of the paper}

As it often happens, the logical presentation of results of a research goes
in opposite direction to their chronological derivation. In 
Section~\ref{sec:BQL+QQL} we collect some geometric 
results from the theory of the B-quadrilateral lattices (BQLs) 
and of quadrilateral
lattices in quadrics (QQLs). Some new results concerning the relation between
(Steiner's) cross-ratios of vertices of elementary
guadrilaterals of elementary hexahedrons of
the QQLs are given there as well. Then in Section~\ref{sec:gen-isoth-latt} 
we define generalized
isothermic lattices and discuss their basic properties. In particular, we give
the synthetic-geometry proof of a basic lemma (the half-hexahedron lemma) which
immediately gives the cross-ratio characterization of the lattices. 
We also present some algebraic consequences (some of them known already
\cite{BobSur}) of the system of Moutard equations supplemented by a quadratic
constraint. In Section~\ref{sec:Darboux} we study in more detail the Darboux
transformation of the
generalized isothermic lattices and the corresponding Bianchi
permutability principle. Finally, in Section~\ref{sec:isothermic} we consider
two dimensional generalized isothermic lattices.
In two Appendices we recall necessary information 
concerning the cross-ratio of four points on a conic curve and we perform some
auxilliary calculations.

\section{The B-quadrilateral lattices and the quadrilateral lattices in quadrics} 
\label{sec:BQL+QQL}
It turns out that compatibility of both BQLs and QQLs with the geometric
integrablity scheme follows from certain classical geometric facts. 
We start each section, devoted to a particular lattice, from the corresponding
geometric statement. 
\subsection{The B-quadrilateral lattice \cite{BQL}}
\begin{figure}
\begin{center}
\includegraphics{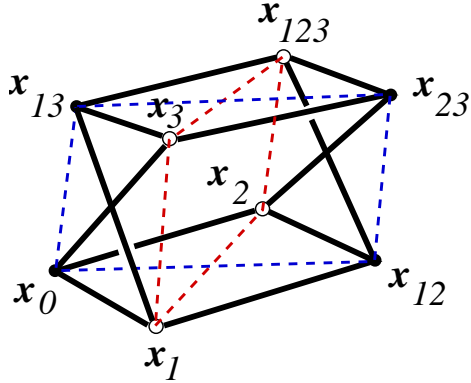}
\end{center}
\caption{Elementary hexahedron of the B-quadrilateral lattice}
\label{fig:moutard}
\end{figure}

\begin{Lem} \label{lem:BKP-hex}
Under hypothesess of Lemma \ref{lem:gen-hex}, assume that the points
$x_0$, $x_{12}$, $x_{13}$, $x_{23}$ are coplanar, then the points
$x_1$, $x_2$, $x_3$, and $x_{123}$ are coplanar as well (see
Figure~\ref{fig:moutard}).
\end{Lem}
As it was discussed in \cite{BQL} the above fact is equivalent to the 
M\"{o}bius theorem (see, for example \cite{Coxeter}) on mutually inscribed
tetrahedra. Another equivalent, but more symmetric,
formulation of Lemma~\ref{lem:BKP-hex} is
provided by the Cox theorem (see \cite{Coxeter}): \emph{Let $\sigma_1$, 
$\sigma_2$, $\sigma_3$, $\sigma_4$ be four planes of general position through a
point $S$. Let $S_{ij}$ be an arbitrary point on the line $\langle \sigma_i,
\sigma_j \rangle$. Let $\sigma_{ijk}$ denote the plane $\langle S_{ij}, S_{ik},
S_{jk} \rangle$. Then the four planes $\sigma_{234}$, $\sigma_{134}$,
$\sigma_{124}$, $\sigma_{123}$ all pass through one point $S_{1234}$.}  
\begin{Def} \label{def:BQL}
A quadrilateral lattice $x:\ZZ^N\to\PP^M$ is called the \emph{B-quadrilateral
lattice} if for any triple of different indices $i,j,k$
the points $x$, $x_{(ij)}$,
$x_{(jk)}$ and $x_{(ik)}$ are coplanar.
\end{Def}
Here and in all the paper, given a fuction $F$ on 
$\ZZ^N$, we denote its shift in the $i$th direction in a standard manner: 
$F_{(i)}(n_1,\dots, n_i, \dots , n_N) = F(n_1,\dots, n_i + 1, \dots , n_N)$.
One can show that a quadrilateral lattice $x:\ZZ^N\to\PP^M$ is a 
B-quadrilateral lattice if and only if
it allows for a homogoneous representation $\bx:\ZZ^N\to\RR^{M+1}_{*}$
satisfying the system of discrete Moutard equations (the
discrete BKP linear problem)
\begin{equation} \label{eq:BKP-lin}
\bx_{(ij)} - \bx = f^{ij} (\bx_{(i)} - \bx_{(j)}) , \quad 1\leq i< j\leq N,
\end{equation}
for suitable functions $f^{ij}:\ZZ^N\to\RR$.

The compatibility condition of the system \eqref{eq:BKP-lin}
implies that the functions $f^{ij}$ can be written
in terms of the potential $\tau:\ZZ^N\to\RR$, 
\begin{equation} \label{eq:tau}
f^{ij} = \frac{\tau_{(i)}\tau_{(j)}}{\tau \, \tau_{(ij)}}, \qquad i\ne j ,
\end{equation}
which satisfies Miwa's discrete
BKP equations \cite{Miwa}
\begin{equation} \label{eq:BKP-nlin}
\tau\, \tau_{(ijk)} = \tau_{(ij)}\tau_{(k)} - \tau_{(ik)}\tau_{(j)} + 
\tau_{(jk)}\tau_{(i)}, \quad 1\leq i< j < k \leq N.
\end{equation}
\begin{Rem}
The trapezoidal lattice \cite{BobSur} is another reduction of the 
quadrilateral lattice being algebraically described by the discrete Moutard
equations \eqref{eq:BKP-lin}. Geometrically, the trapezoidal lattices are
characterized by parallelity of diagonals of the elementary quadrilaterals,
thus they belong to the affine geometry. 
Moreover, because the trapezoidal constraint is imposed on the level
of elementary quadrilaterals then,
from the point of view of the geometric
integrability scheme, one has to check its three
dimensional consistency. In contrary, the BQL constraint is imposed on the level
of elementary hexahedrons, and to prove geometrically its integrability
one has to check four dimensional
consistency. 
\end{Rem}

\subsection{The quadrilateral lattices in quadrics}

\begin{Lem} \label{lem:qred-hex}
Under hypotheses of Lemma \ref{lem:gen-hex}, assume that the points
$x_0$, $x_1$, $x_2$, $x_3$, $x_{12}$, $x_{13}$, $x_{23}$ belong to a 
quadric $\mathcal{Q}$. Then the point $x_{123}$ belongs to 
the quadric $\mathcal{Q}$ as well.
\end{Lem}
\begin{Rem}
The above fact is a consequence of the classical 
\emph{eight points theorem} (see, for example \cite{Coxeter}) which says that
\emph{seven points in general position determine a unique eighth point, such
that every quadric through the seven passes also through the eighth}. In our
case the point $x_{123}$ is contained in the three (degenerate) quadrics being
pairs of opposite facets of the hexahedron. 
\end{Rem}
\begin{Rem}
Lemma \ref{lem:BKP-hex} can be considered as a "reduced" resion of 
Lemma~\ref{lem:qred-hex} when the quadric $\mathcal{Q}$
degenerates to a pair of planes.
\end{Rem}
\begin{Def} \label{def:QQL}
A quadrilateral lattice $x:\ZZ^N\to\mathcal{Q}\subset\PP^M$ 
contained in a hyperquadric $\mathcal{Q}$ is called the 
\emph{$\mathcal{Q}$-reduced quadrilateral lattice}  (QQL).
\end{Def}
Integrability of the QQLs was pointed out in \cite{q-red}, where also the
corresponding Darboux-type transformation (called in this context the Ribaucour
transformation) was constructed in the vectorial form.   
When the quadric is irreducible then generically 
it cuts the planes of the hexahedron along
conics. 
\begin{Def}
A quadrilateral lattice $x:\ZZ^N\to\mathcal{Q}\subset\PP^M$ in a
hyperquadric $\mathcal{Q}$ such that the intersection of the planes of
elementary quadrilaterals of the latice with the quadric
are irreducible conic curves is called
\emph{locally irreducible}.
\end{Def}
The following result, which will not be used in the sequel and
whose proof can be found in Appendix~\ref{sec:aux-calc}, 
generalizes 
the relation between complex cross-ratios of the opposite quadrilaterals of
elementary hexahedrons of the circular lattices \cite{Bobenko-O}.
\begin{Prop} \label{prop:qred-cr}
Given locally irreducible quadrilateral lattice 
$x:\ZZ^N\to\mathcal{Q}\subset\PP^M$ in a
hyperquadric $\mathcal{Q}$, denote by
\begin{equation*}
\lambda^{ij} = \qcr(x_{(i)},x_{(j)};x,x_{(ij)}), \qquad 1\leq i<j\leq N,
\end{equation*} 
the cross-ratios (defined with respect to 
the corresponding conic curves) 
of the four vertices of the quadrilaterals. 
Then the cross-ratios are related by the following system of equations
\begin{equation} \label{eq:qred-cross-r}
\lambda^{ij}\lambda^{ij}_{(k)}\lambda^{jk}\lambda^{jk}_{(i)} =
\lambda^{ik}\lambda^{ik}_{(j)}, \quad 1\leq i<j<k\leq N.
\end{equation}
\end{Prop}
\begin{Rem}
The system \eqref{eq:qred-cross-r} can be considered as the gauge invariant
integrable difference equation governing QQLs.
\end{Rem}

\section{Generalized isothermic lattices}
\label{sec:gen-isoth-latt}
Because simultaneous application of integrable constraints preserves
integrability we know \emph{a priori} that the following reduction of the
quadrilateral lattice is integrable.
\begin{Def}
A B-quadrilateral lattice in a hyperquadric $x:\ZZ^N\to\mathcal{Q}\subset\PP^M$
satisfying the local irreducibility condition is called a 
\emph{generalized isothermic lattice}.
\end{Def}
\subsection{The half hexahedron lemma and its consequences}
We start again from a
geometric result, which leads to the cross-ratio characterization 
of the generalized
isothermic lattice.
\begin{Lem}[The half hexahedron lemma] \label{lem:half-hex}
Under hypotheses of Lemmas \ref{lem:BKP-hex} and \ref{lem:qred-hex} and
assuming irreducibility of the conics of the intersection of the 
corresponding planes with the quadric
we have
\begin{equation} \label{eq:product-cr}
\qcr(x_1,x_3;x_0,x_{13}) = \qcr(x_1,x_2;x_0,x_{12}) \,
\qcr(x_2,x_3;x_0,x_{23}),
\end{equation}
where the cross-ratios are defined with respect to the corresponding
conics. 
\end{Lem} 

\begin{proof}
Denote (see
Figure~\ref{fig:hex-lem}) 
\begin{figure}
\begin{center}
\includegraphics{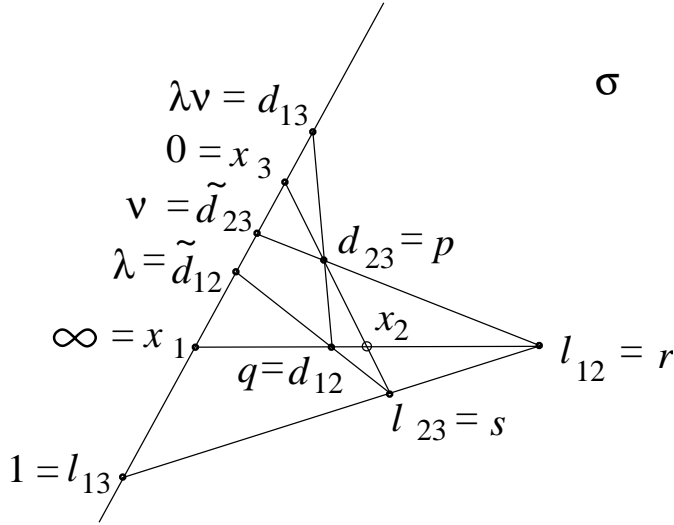}
\end{center}
\caption{Geometric proof of Lemma \ref{lem:half-hex}}
\label{fig:hex-lem}
\end{figure}the plane $\sigma=\langle x_1, x_2, x_3 \rangle$, and 
represent points of the conic
$\mathcal{C}_{ij}=\cQ\cap\langle x_0, x_i, x_j \rangle$, $1\leq i < j \leq 3$
by points of the 
line $\langle x_i, x_j \rangle$, via the corresponding
planar pencil with the base at $x_0$. In this way the projective structure of
the conics conicides with that of the corresponding lines.

By $\ell_{ij}$ denote the
intersection of the tangent line to $\mathcal{C}_{ij}$ at $x_0$ with $\sigma$.
Notice that the points $\ell_{ij}$ belong to
the intersection line of the tangent plane to the quadric at $x_0$ 
with $\sigma$, and all they represent 
$x_0$ but from the point of view of different conics.
Denote
by $d_{ij}$ the intersection point of the line
$\langle x_0, x_{ij} \rangle$ with the plane $\sigma$. Notice that coplanarity 
of the points 
$x_0$, $x_{12}$, $x_{23}$ and  $x_{23}$ is equivalent to collinearity
of $d_{12}$, $d_{23}$ and $d_{13}$. Moreover, by definition of
the cross-ratio on conics, we have
\begin{equation} \label{eq:cr-c-l}
\qcr(x_i,x_j;x_0,x_{ij})  = \qcr(x_i,x_j;\ell_{ij},d_{ij}).
\end{equation}

To find a relation between the cross-ratios  
consider perspectivity between the lines $\langle x_1, x_2 \rangle$ and
$\langle x_1, x_3 \rangle$ with the center $\ell_{23}$. It transforms 
$x_2$ into $x_3$, $x_1$ into $x_1$, $d_{12}$ into $\tilde{d}_{12}$ (this is
just definition of $\tilde{d}_{12}$) and $\ell_{12}$ into $\ell_{13}$, 
therefore
\begin{equation}
\qcr(x_1,x_2;\ell_{12},d_{12}) = 
\qcr(x_1,x_3;\ell_{13},\tilde{d}_{12}).
\end{equation}
Smilarly, considering perspectivity between the lines 
$\langle x_2, x_3 \rangle$ and
$\langle x_1, x_3 \rangle$ with the center $\ell_{12}$ we obtain
\begin{equation}
\qcr(x_2,x_3;\ell_{23},d_{23}) = 
\qcr(x_1,x_3;\ell_{13},\tilde{d}_{23}),
\end{equation}
where again $\tilde{d}_{23}$ is the projection of $d_{23}$. 
The comparison of Figures \ref{fig:hex-lem} and \ref{fig:multiplication} 
gives
\begin{equation} \label{eq:cr-l-pr}
\qcr(x_1,x_3;\ell_{13},\tilde{d}_{12})
\qcr(x_1,x_3;\ell_{13},\tilde{d}_{23})=
\qcr(x_1,x_3;\ell_{13},d_{13}),
\end{equation}
which because of equations \eqref{eq:cr-c-l}-\eqref{eq:cr-l-pr} implies the
statement.
\end{proof}
\begin{Rem}
For those who do not like synthetic geometry proofs we give the algebraic proof
of the above Lemma in Appendix~\ref{sec:aux-calc}.
\end{Rem}
\begin{Cor}
Equation \eqref{eq:product-cr} can be written in a more symmetric form
\begin{equation} \label{eq:product-cr-symm}
\qcr(x_1,x_2;x_0,x_{12}) \,
\qcr(x_2,x_3;x_0,x_{23}) \, \qcr(x_3,x_1;x_0,x_{13})  = 1.
\end{equation}
\end{Cor}
\begin{Cor}
The cross-ratio of the four (coplanar) points $x_0$, $x_{12}$, $x_{13}$ and 
$x_{23}$ can be expressed by the other cross-ratios as
\begin{equation}
\qcr(x_0,x_{12};x_{13},x_{23})=\qcr(x_0,x_{2};x_{3},x_{23})
\qcr(x_1,x_{0};x_{3},x_{13}).
\end{equation}
\end{Cor}
\begin{proof}
Consider the line $\langle d_{13}, d_{23} \rangle$, which is the section of the
planar pencil containing lines $\langle x_0, x_{13} \rangle$ and
$\langle x_0, x_{23} \rangle$ with the plane $\sigma$. Denote by $\ell$ the
intersection point of the line with the line 
$\langle \ell_{13}, \ell_{23} \rangle$, then
\begin{equation*}
\qcr(x_0,x_{12};x_{13},x_{23}) = \qcr(\ell,d_{12};d_{13},d_{23}).
\end{equation*}
After projection from $\ell_{12}$ we have, in notation of 
Figure~\ref{fig:hex-lem}, 
\begin{equation*}
\qcr(\ell,d_{12};d_{13},d_{23}) = \qcr(\ell_{13},x_1;d_{13},\tilde{d}_{23})=
\qcr(1,\infty;\lambda\nu,\nu).
\end{equation*}
Then the standard permutation properties of the cross-ratio give the statement. 
\end{proof}

\begin{figure}
\begin{center}
\includegraphics{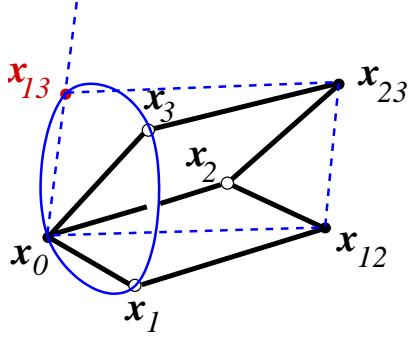}
\end{center}
\caption{Construction of the point $x_{13}$ from points $x_0$, $x_{1}$, $x_{2}$, 
$x_{3}$, $x_{12}$ and $x_{23}$. It belongs to the
intersection line of two planes $\langle x_0, x_{1}, x_{3} \rangle$ with the
plane $\langle x_0, x_{12}, x_{23} \rangle$. Because the line intersects the
quadric at $x_0$, it must have also the second intersection point.}
\label{fig:twotothree}
\end{figure}

\begin{Cor}[The hexahedron lemma] \label{cor:hex}
Under assumption of Lemma \ref{lem:half-hex} the cross-ratios on opposite
quadrilaterals of the hexahedron are equal, i.e.
\begin{align}
\nonumber
\qcr(x_1,x_2;x_0,x_{12}) & = \qcr(x_{13},x_{23};x_3,x_{123}) , \\
\label{eq:equalities-cr}
\qcr(x_2,x_3;x_0,x_{23}) & = \qcr(x_{12},x_{13};x_1,x_{123}) , \\
\nonumber
\qcr(x_1,x_3;x_0,x_{13}) & = \qcr(x_{12},x_{23};x_2,x_{123}) . 
\end{align}
\end{Cor}
\begin{proof}
Equation \eqref{eq:product-cr-symm} written for the three quadrilaterals
meeting in $x_3$ reads
\begin{equation} 
\qcr(x_0,x_{12};x_3,x_{1}) \,
\qcr(x_{13},x_{23};x_3,x_{123}) \, \qcr(x_{23},x_0;x_3,x_{2})  = 1,
\end{equation}
which compared with \eqref{eq:product-cr-symm} gives, after using elementary
properties of the cross-ratio, the first equation of
\eqref{eq:equalities-cr}. Others can be obtained similarly.
\end{proof}
\begin{Cor}
By symmetry we have also
\begin{equation}
\qcr(x_0,x_{12};x_{13},x_{23}) =  \qcr(x_{1},x_{2};x_{3},x_{123}).
\end{equation}
\end{Cor}
\begin{Rem}
Notice that two neighbouring facets of the above hexahedron determine the
whole hexahedron via construction visualized on Fig.~\ref{fig:twotothree}. 
\end{Rem}
\begin{Rem}
It is easy to see that, unlike in the case of isothermic lattices,
three vertices of a quadrilateral of trapezoidal lattice in a quadric
\cite{BobSur} determine
the forth vertex.
\end{Rem}

\begin{Prop} \label{prop:cross-ratio-gen}
A quadrilateral lattice in a quadric $x:\ZZ^N\to\mathcal{Q}\subset\PP^M$ 
satisfying the local irreducibility condition is a generalized isothermic
lattice if and only if
there exist functions $\alpha^{i}:\ZZ\to\RR$ of single arguments $n_i$ such
that the cross-ratios $\lambda^{ij} = \qcr(x_{(i)},x_{(j)};x,x_{(ij)})$ 
can be factorized as follows
\begin{equation} \label{eq:isot-cross-res}
\lambda^{ij} = \frac{\alpha^i}{\alpha^j}, \qquad 1\leq i<j\leq N.
\end{equation}
\end{Prop}
\begin{proof}
Equations \eqref{eq:product-cr} and \eqref{eq:equalities-cr} can be rewritten as
\begin{equation} \label{eq:izot-cross-r-1}
\lambda^{ij}\lambda^{jk} = \lambda^{ik}, \quad 1\leq i<j<k\leq N,
\end{equation}
and
\begin{equation} \label{eq:izot-cross-r-op}
\lambda^{ij}=\lambda^{ij}_{(k)}, \quad \lambda^{jk} = \lambda^{jk}_{(i)}, \quad
\lambda^{ik}=\lambda^{ik}_{(j)}, \quad 1\leq i<j<k\leq N;
\end{equation}
notice their consistency with
the general system \eqref{eq:qred-cross-r}.

Equations \eqref{eq:izot-cross-r-1}-\eqref{eq:izot-cross-r-op} imply that
cross-ratios of two dimensional sub-lattices of the
generalized isothermic lattice satisfy condition of the form 
\eqref{eq:isoth-cr}, i.e.,
\begin{equation}
\lambda^{ij}_{(ij)}\lambda^{ij} = \lambda^{ij}_{(i)}\lambda^{ij}_{(j)},
\quad 1\leq i<j\leq N.
\end{equation}
For a fixed pair $i,j$, the above relation can be resolved as in 
\eqref{eq:isot-cross-res} (the first equation in \eqref{eq:izot-cross-r-op}
asserts that the functions $\alpha^i$ and $\alpha^j$ are the same for all $i,j$
sublattices). Finally, equations \eqref{eq:izot-cross-r-1} imply the 
the functions $\alpha$ can be defined consistently on the whole lattice.
\end{proof}
For convenience of the Reader we present also 
the algebraic proof of the above
properties of the generalized isothermic lattice
(see also \cite{BobSur} for analogous results concerning T-nets in a quadric). 
\begin{proof}[The algebraic proof]
Assume that solutions of the system of the discrete Moutard equations
\eqref{eq:BKP-lin} satisfy the quadratic constraint
\begin{equation} \label{eq:q-constr}
(\bx | \bx) = 0,
\end{equation} 
where $(\cdot | \cdot ) $ is a symmetric nondegenerate bilinear form. Then the
coefficients of the Moutard equations should be of the form
\begin{equation} \label{eq:f-isoth}
f^{ij} = \frac{(\bx | \bx_{(i)} - \bx_{(j)})}{(\bx_{(i)} | \bx_{(j)})},
\quad 1\leq i<j\leq N.
\end{equation}
Moreover by direct calculations one shows that
\begin{equation}
(\bx_{(i)} | \bx)_{(j)} = (\bx_{(i)} | \bx),
\quad 1\leq i<j\leq N,
\end{equation}
which implies that the products
$(\bx_{(i)} | \bx) $, which we denote by $\alpha_i$, 
are functions of single variables $n_i$.

Consider the points
$x, x_{(i)}, x_{(j)}$ as the (projective) basis of 
the plane $\langle x, x_{(i)}, x_{(j)} \rangle$. 
Then the homogeneous coordinates of points of the plane can be
written as
\begin{equation}
\by = t \bx + t_i \bx_{(i)} + t_j \bx_{(j)}, \qquad (t,t_i,t_j) \in \RR^3_*, 
\end{equation}
modulo the standard common proportionality factor. In particular, the line
$\langle x, x_{(i)} \rangle$ is given by equation $t_j = 0$, 
and the line
$\langle x, x_{(j)} \rangle$ is given by equation $t_i = 0$. 
Due to the discrete Moutard equation \eqref{eq:BKP-lin}
the line $\langle x, x_{(ij)} \rangle$ is given by equation $t_i + t_j = 0$.

To find the cross-ratio $ \qcr(x_{(i)},x_{(j)};x,x_{(ij)})=\lambda^{ij} $ via
lines of the planar pencil with the base point $x$ we need equation of
the tangent
to the conic $(\by | \by ) =0$ at that point.
It is easy to check that the conic is given by
\begin{equation}
t t_i \alpha_{i} + t t_j \alpha_{j} + t_i t_j  (\bx_{(i)} | \bx_{(j)}) =0.
\end{equation}
The tangent to the conic at $x$ is then given by
\begin{equation}
t_i \alpha_{i} + t_j \alpha_{j} =0, 
\end{equation}
which implies equation \eqref{eq:isot-cross-res}.
\end{proof}
\begin{Rem}
It should be mentioned the a similar quadratic
reduction of the discrete Moutard 
equation (for $N=2$) appeared in a paper of Wolfgang Schief~\cite{Schief-C} 
under the name
of discrete vectorial
Calapso equation, as an integrable discrete vectorial analogue of the 
Calapso equation~\cite{Calapso}, which is of the fourth order and describes 
isothermic surfaces. It turns out that the discrete Calapso equation
describes also the so called Bianchi reduction of discrete asymptotic 
surfaces~\cite{DNS-Bianchi-ass}.
\end{Rem}

\subsection{Isothermic lattices in the M\"{o}bius sphere
and the so called
Clifford configuration}
\begin{figure}
\begin{center}
\includegraphics[width=8cm]{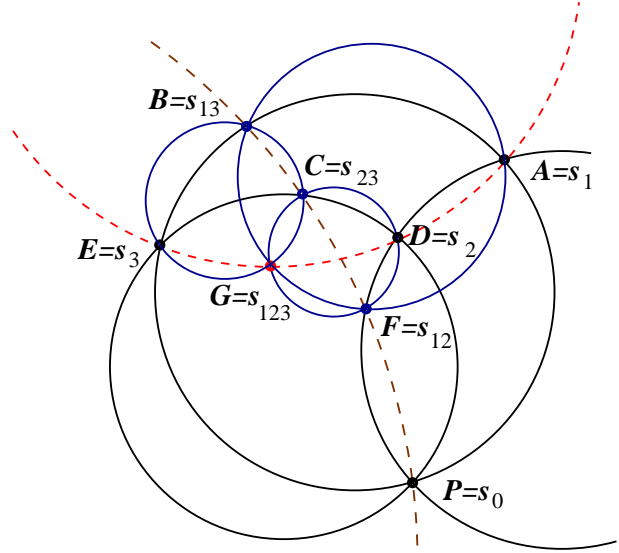}
\end{center}
\caption{The so called
Clifford configuration of circles (the second Miquel configuration)}
\label{fig:Clifford}
\end{figure}

In \cite{KoSch-Clifford} Konopelchenko and Schief 
observed that the "complex cross-ratio
definition" of the discrete isothermic surfaces, when extended to three
dimensional lattices, is related with the so called
Clifford configuration of circles (see Figure~\ref{fig:Clifford}). 
In this Section we would like to explain that fact geometrically. 

\begin{figure}
\begin{center}
\includegraphics[width=8cm]{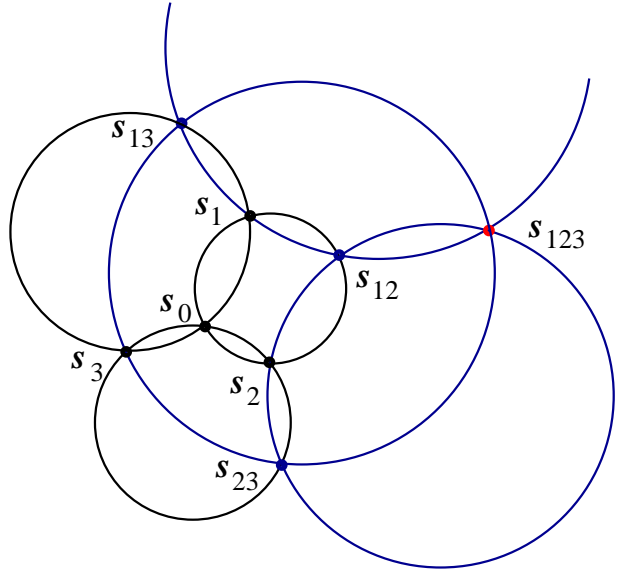}
\end{center}
\caption{The (first) Miquel configuration of circles}
\label{fig:Miquel}
\end{figure}

Our point of view on the Clifford configuration is closely related to the
geometric definition of the isothermic lattice in the M\"{o}bius sphere.
Therefore we start with another, less restrictive, 
configuration of circles on the plane, the Miquel configuration
(see Figure~\ref{fig:Miquel}), which 
provides geometric explanation of integrability of the circular 
lattice~\cite{CDS}. When the quadric in Lemma \ref{lem:qred-hex} is the 
standard sphere
then the intersection curves of the planes of the quadrilaterals with 
the sphere
are circles. After the stereographic projection from a generic point of the 
sphere we obtain the classical Miquel theorem \cite{Miquel}, which can
be stated as follows (given three distinct points $a$, $b$ and $c$, 
by $C(a,b,c)$ we denote the unique circle-line passing through them).

\begin{Th}[The Miquel configuration] \label{th:Miquel}
Given four coplanar points $s_0$, $s_i$, $i=1,2,3$. On each circle 
$C(s_0,s_i, s_j)$, $1\leq i < j \leq 3$ choose a point, denoted correspondingly 
by $s_{ij}$. Then there exists the unique point $s_{123}$ which belongs
simultaneously to the three circles  $C(s_1, s_{12}, s_{13})$, 
$C(s_2, s_{12}, s_{23})$ and $C(s_3, s_{13}, s_{23})$.
\end{Th}
The additional assumption about coplanarity of the points 
$x_0$, $x_{12}$, $x_{13}$,
$x_{23}$ on the sphere is then equivalent to the 
additional assumption about concircularity of the corresponding points 
$s_0$, $s_{12}$, $s_{13}$, $s_{23}$. In view of Lemma \ref{lem:BKP-hex}
we obtain therefore another configuration
of circles, the so called
Clifford configuration, which can be described as follows.

\begin{Th}[The so called Clifford configuration] \label{th:Clifford}
Under hypotheses of Theorem \ref{th:Miquel} assume that the points
$s_0$, $s_{12}$, $s_{13}$, $s_{23}$ are concircular, then the points 
$s_1$, $s_{2}$, $s_{3}$ and $s_{123}$ are concircular as well.
\end{Th} 
\begin{Rem}
The original Clifford's formulation of the above result was more symmetric. Its
relation to Theorem~\ref{th:Clifford} is analogous to relation of the Cox
theorem to Lemma~\ref{lem:BKP-hex}. We remark that although
the above theorem is usually attributed (see for example \cite{Coxeter})
to Wiliam Clifford \cite{Clifford} it appeared 
in much earlier paper \cite{Miquel} of Auguste Miquel, where we read as
\emph{Th\`{e}or\`{e}me II} the following statement: 
\emph{Lorsqu'un quadrilat\`{e}re complet curviligne ABCDEF
est form\'{e} par quatre arcs de cercle AB, BC, CD, DA, qui se coupent tous
quatre en un m\^{e}me point P, si l'on circonscrit des circonf\'{e}rences de
cercle \`{a} chacun des quatre triangles curvilignes que forment les
c\^{o}t\'{e}s de ce quadrilat\`{e}re, les circonf\'{e}rences de cercle AFB, EBC,
DCF, DAE ainsi obtenues se couperont toutes quatre en un m\^{e}me point G}.
Points on Figure~\ref{fig:Clifford} are labelled in double way
to visualize simultaneously
the configuration in formulation of Theorem~\ref{th:Clifford} and in Miquel's
formulation.  
\end{Rem}

\section{The Darboux transformation of the generalized isotermic lattice}
\label{sec:Darboux} 
\subsection{The fundamental, Moutard and Ribaucour transformations}
Usually, on the discrete level there is no essential difference between 
integrable lattices and their transformations. 
The analogue of the fundamental transformation of Jonas for
quadrilateral lattices is defined as construction of a new level of the lattice
\cite{TQL} keeping the basic property of planarity of elementary quadrilaterals.
Below we recall the relevant definitions of the fundamental transformation and
its important reductions -- the BQL reduction \cite{BQL}
(algebraically equivalent to the Moutard transformation \cite{NiSchief}), 
and the QQL reduction \cite{q-red}
called the Ribaucour transformation.

\begin{Def}
\emph{The fundamental transform} of a quadrilateral lattice 
$x:\ZZ^N\to\PP^M$ is a new quadrilateral lattice
$\hat{x}:\ZZ^N\to\PP^M$ constructed under assumption that for any point
$x$ of the lattice and any direction $i$, the four points
$x$, $x_{(i)}$, $\hat{x}$ and $\hat{x}_{(i)}$ are coplanar.
\end{Def}
\begin{Def}
The fundamental transformation of a B-quadrilateral lattice 
$x:\ZZ^N\to\PP^M$  
constructed under additional assumption that for any point
$x$ of the lattice and any pair $i,j$ of different directions, the four points
$x$, $x_{(ij)}$, $\hat{x}_{(i)}$ and $\hat{x}_{(j)}$ are coplanar is called
\emph{the BQL (Moutard) reduction} of the fundamental transformation of $x$.
\end{Def}
Algebraic description of the above transformation is given as follows 
\cite{NiSchief}.
Given solution $\bx$ of the system of discrete Moutard equations
\eqref{eq:BKP-lin} and given its scalar solution $\theta$, then
the solution $\hat{\bx}$ of the system 
\begin{equation} \label{eq:Moutard-transf}
\hat\bx_{(i)} - \bx = \frac{\theta}{\theta_{(i)}}\left( \hat\bx -
\bx_{(i)}\right),
\end{equation}
satisfies equations \eqref{eq:BKP-lin} with the new potential
\begin{equation}
\hat{f}^{ij} = f^{ij}\frac{\theta_{(i)}\theta_{(j)}}{\theta\theta_{(ij)}},
\qquad i<j,
\end{equation}
and new $\tau$-function
\begin{equation} \label{eq:transf-Mout-tau}
\hat\tau = \theta\tau.
\end{equation}

\begin{Def}
The fundamental transformation of a quadrilateral lattice 
$x:\ZZ^N\to\mathcal{Q}\subset\PP^M$ in a quadric, 
constructed under additional assumption
that also $\hat{x}$ satifies the same quadratic constraint
is called
\emph{the Riboucour transformation} of $x$.
\end{Def}

\subsection{The Darboux transformation}

\begin{Def}
The fundamental transformation of a generalized isothermic lattice which is
simultaneously the Ribaucour and the Moutard transformation is called 
\emph{the Darboux transformation}.
\end{Def}
Notice that there is essentially no difference between
the Darboux transformation and  construction of a new
level of the generalized isothermic lattice. Therefore, given points $x$,
$x_{(i)}$, $x_{(j)}$, $x_{(ij)}$, $i\ne j$, of the initial lattice, and 
given points $\hat{x}$, $\hat{x}_{(j)}$ of its Darboux transform then the point
$\hat{x}_{(i)}$ is determined by the "half-hexahedron construction" visualized
on Figure~\ref{fig:twotothree}, i.e., $\hat{x}_{(i)}$ is the intersection point
of the line 
$\langle x, x_{(i)}, \hat{x} \rangle \cap 
\langle x, x_{(ij)}, \hat{x}_{(j)} \rangle$ with the quadric. Moreover, 
Lemma~\ref{lem:half-hex} implies
\begin{equation} \label{eq:cr-prod-transf}
\qcr(x_{(i)},\hat{x};x,\hat{x}_{(i)}) = \qcr(x_{(i)},x_{(j)};x,x_{(ij)})
\qcr(x_{(j)},\hat{x};x,\hat{x}_{(j)}),
\end{equation}
while Corollary~\ref{cor:hex} gives
\begin{equation}
\qcr(\hat{x}_{(i)},\hat{x}_{(j)};\hat{x},\hat{x}_{(ij)})=
\qcr(x_{(i)},x_{(j)};x,x_{(ij)}).
\end{equation}

The algebraic derivation of the above results is given below.
The Darboux
transformations of the discrete isothermic surfaces in the light-cone
description were discussed in a similar spirit in \cite{BobSur}.
\begin{Prop} \label{prop:Darboux-prod}
If $\hat{x}:\ZZ^N\to\mathcal{Q}\subset\PP^M$ is a Darboux transform of the 
generalized isothermic lattice $x:\ZZ^N\to\mathcal{Q}\subset\PP^M$ then
the product $(\hat{\bx}|\bx)$ of their homogeneous coordinates in the gauge
of the linear problem \eqref{eq:BKP-lin} and the corresponding Moutard
transformation \eqref{eq:Moutard-transf}, with respect to the bilinear form
$(\cdot|\cdot)$ defining the quadric $\mathcal{Q}$, is constant.
\end{Prop}
\begin{proof}
The homogeneous coordinates $\bx$ and $\hat{\bx}$ in the Moutard 
transformation satisfy equation of the form
\begin{equation}
\hat{\bx}_{(i)} - \bx = f^i(\hat{\bx} - \bx_{(i)}), \qquad 1\leq i \leq N,
\end{equation}
with appropriate functions $f^i:\ZZ^N\to\RR$. The quadratic condition
$(\hat{\bx}_{(i)}|\hat{\bx}_{(i)})=0$ together with other quadratic conditions
give
\begin{equation} \label{eq:fi-isoth}
f^i = \frac{(\bx| \hat{\bx} -\bx_{(i)})}{(\hat{\bx}|\bx_{(i)})},
\end{equation}
which implies $(\hat{\bx}_{(i)}|{\bx}_{(i)}) = (\hat{\bx}|\bx)$.
\end{proof}
Notice the above proof goes along the corresponding reasoning in the first part of
the algebraic proof of Proposition \ref{prop:cross-ratio-gen}. The analogous
reasoning as in its second part gives the following statement.
\begin{Cor}
Under hypothesis of Proposition \ref{prop:Darboux-prod} denote 
$\zeta=(\hat{\bx}|\bx)$ and $\lambda^i = \qcr(\hat{x},x_{(i)};x,\hat{x}_{(i)})$
then equations \eqref{eq:isot-cross-res} and \eqref{eq:izot-cross-r-1}
should be replaced by
\begin{equation} \label{eq:cr-lambda-prod}
\lambda^i = \frac{\zeta}{\alpha_i}, \qquad \lambda^i \lambda^{ij} = \lambda^j. 
\end{equation}
\end{Cor}
The above reasoning can be reversed giving the algebraic way to find the
Darboux transform of a given generalized isothermic lattice. 
\begin{Th} \label{th:Darboux}
Given a solution $\bx :\ZZ^N\to \RR^{M+1}_*$ of the system of Moutard equations
\eqref{eq:BKP-lin} satisfying the constraint $(\bx|\bx)=0$, considered as
homogeneous coordinates of generalized isothermic lattice 
$x:\ZZ^N\to\mathcal{Q}\subset\PP^M$, denote $\alpha_i = (\bx_{(i)}|\bx)$. 
Given a
point $[\hat{\bx}_0]=\hat{x}_0\in\mathcal{Q}$, denote $\zeta =
(\hat{\bx}_0|\bx(0))$. Then there exists unique solution of the linear system
\begin{equation} \label{eq:Darboux}
\hat{\bx}_{(i)} = \bx + \frac{\zeta - \alpha_i}{(\hat{\bx}|\bx_{(i)})}
(\hat{\bx} - \bx_{(i)}), \qquad, \quad 1\leq i \leq N,
\end{equation} 
with initial condition $\hat{\bx}(0) = \hat{\bx}_0$ which gives the Darboux
transform of the lattice $x$. In particular
\begin{align} \label{eq:cr-zeta-alpha}
\qcr(\hat{x},x_{(i)};x,\hat{x}_{(i)}) & = \frac{\zeta}{\alpha_i} \\
\qcr(\hat{x}_{(i)},\hat{x}_{(j)};\hat{x},\hat{x}_{(ij)}) & = 
\qcr(x_{(i)},x_{(j)};x,x_{(ij)}) =
\frac{\alpha^i}{\alpha^j}.
\end{align}
\end{Th}
Before proving the Theorem let us state a Lemma
relating the parameter $\zeta$ of the Darboux transformation with the functional
parameter $\theta$ of the Moutard transformation \eqref{eq:Moutard-transf}.
\begin{Lem} \label{eq:zeta-theta}
Under hypotheses of Theorem \ref{th:Darboux} the solution $\theta$ of the system
\begin{equation}
\theta_{(i)}=\theta\frac{(\hat{\bx}|\bx_{(i)})}{\zeta - \alpha_i}, \qquad
\quad 1\leq i \leq N,
\end{equation}
satisfies the system of discrete Moutard equations \eqref{eq:BKP-lin}
with the coefficients given by \eqref{eq:f-isoth}.
\end{Lem}
\begin{proof}[Proof of the Lemma]
By direct verification. Notice that both ways to calculate
$\theta_{(ij)}$, $i\ne j$, from $\theta$ give the same result, and to do that we
do not use compatibility of the system \eqref{eq:Darboux}.
\end{proof}
\begin{proof}[Proof of the Theorem]
By direct calculation one can check that the system \eqref{eq:Darboux} 
preserves
the constraints $(\hat{\bx}|\hat{\bx}) =0$ and $(\hat{\bx}|\bx)=\zeta$,
moreover
$(\hat{\bx}_{(i)}|\hat{\bx}) =(\bx_{(i)}|\bx)$.
Compatibility of the system \eqref{eq:Darboux} can be checked by direct
calculation, but in fact it is the consequence of 
Lemma \ref{eq:zeta-theta} and properties of the Moutard 
transformation \eqref{eq:Moutard-transf}.
\end{proof}
\begin{Rem}
Notice that because there is essentially no diffrence between the lattice
directions and the transformation directions, the tranformation equations
\eqref{eq:Darboux} can be guessed by keeping the Moutard-like form supplementing
it by calculation of the coefficient $f^i$ from the quadratic constraint.
We will use this observation in the next Section where we consider the
permutability principle for the Darboux transformations of generalized 
isothermic lattices.
\end{Rem}
\subsection{The Bianchi permutability principle}
The original Bianchi superposition principle for the Darboux transformations of
the isothermic surfaces reads as follows \cite{Bianchi-isoth}:
\emph{Se dalla superficie isoterma $S$ si ottengono due nuove superficie
isoterme $S_1$, $S_2$ mediante le trasformazioni di Darboux $D_{m_1}$, $D_{m_2}$ a
costanti $m_1$, $m_2$ differenti, esiste una quarta superficie isoterma
$\overline{S}$, pienamente determinata e costruibile in termini finiti, che
\`{e} legata alla sua volta alle medesime superficie $S_1$, $S_2$ da due
trasformazioni di Darboux $\overline{D}_{m_2}$, $\overline{D}_{m_1}$ colle
costanti invertite $m_2$, $m_1$.}
\begin{figure}
\begin{center}
\includegraphics{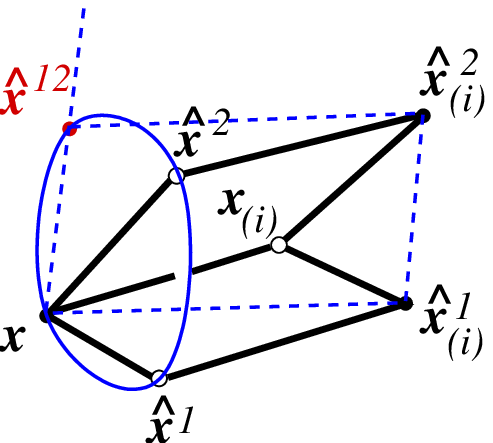}
\end{center}
\caption{Geometric construction of the superposition of two Darboux
transformations}
\label{fig:bianchi}
\end{figure}
Its version for generalized isothermic lattices can be formulated analogously.
\begin{Prop} \label{prop:superp}
When from given generalized isothermic lattice $x$ there were constructed two
new isothermic lattices $\hat{x}^1$ and $\hat{x}^2$ via the Darboux transformations with
different parameters $\zeta_1$ and $\zeta_2$, then there exists the unique forth
generalized isothermic lattice $\hat{x}^{12}$, determined in algebraic terms from the
three previous ones, which is connected with two intermediate lattices 
$\hat{x}^1$ and
$\hat{x}^2$ via two Darboux transformations with reversed parameters $\zeta_2$,
$\zeta_1$. 
\end{Prop}
\begin{proof}
The algebraic properties of the B-reduction of the fundamental transformation
(the discrete Moutard transformation) imply that in 
the gauge of the linear problem \eqref{eq:BKP-lin} and of the transformation
equations \eqref{eq:Moutard-transf} the superposition of two such
transformations reads
\begin{equation} \label{eq:-transf-sup}
\hat{\bx}^{12} - \bx = f (\hat{\bx}^1 - \hat{\bx}^2),
\end{equation}
where $f$ is an appropriate function \cite{NiSchief,BQL}. Because of the
additional quadratic constraints the function is given by (compare also equations 
\eqref{eq:f-isoth} and \eqref{eq:f-isoth})
\begin{equation} \label{eq:f-isoth-sup}
f = \frac{(\bx | \hat{\bx}^1 - \hat{\bx}^2)}{(\hat{\bx}^1 | \hat{\bx}^2)}.
\end{equation} 
The lattice $\hat{x}^{12}$ with homogeneous coordinates
given by \eqref{eq:-transf-sup} and \eqref{eq:f-isoth-sup} is superposition
of two Darboux transforms. Finally, direct calculation shows that
\begin{equation}
(\hat{\bx}^{12} | \hat{\bx}^2) =
(\hat{\bx}^1 | \bx) = \zeta_1, \qquad
(\hat{\bx}^{12} | \hat{\bx}^1)=
(\hat{\bx}^2 | \bx)= \zeta_2 .
\end{equation}
\end{proof}
\begin{Cor}
The final algebraic superposition formula reads
\begin{equation}
\hat{\bx}^{12} - \bx = \frac{\zeta_1 - \zeta_2}{(\hat{\bx}^1 | \hat{\bx}^2)}
(\hat{\bx}^1 - \hat{\bx}^2),
\end{equation}
while the cross-ratio of the four corresponding points calculated with respect
to the conic intersection of the plane $\langle x,\hat{x}^1,\hat{x}^2 \rangle$ 
and the quadric is given by
\begin{equation}\label{eq:cr-zeta-sup}
 \qcr(\hat{x}^1,\hat{x}^2;x,\hat{x}^{12}) = \frac{\zeta_1}{\zeta_2}.
\end{equation}
\end{Cor}
To find the lattice $\hat{x}^{12}$ geometrically we can use again
the "half-hexahedron construction" in the new context 
visualized on Figure~\ref{fig:bianchi}
(compare with Figure~\ref{fig:twotothree}).
Moreover, Lemma~\ref{lem:half-hex} gives
\begin{equation} \label{eq:cr-prod-sup}
\qcr(\hat{x}^1,\hat{x}^2;x,\hat{x}^{12}) =
\qcr(\hat{x}^1, x_{(i)};x,\hat{x}^1_{(i)}) \,
\qcr(x_{(i)}, \hat{x}^2;x,\hat{x}^2_{(i)}),
\end{equation}
which, due to equation \eqref{eq:cr-zeta-alpha}, is in agreement with
\eqref{eq:cr-zeta-sup}.

\section{Two dimensional generalized isothermic lattice}
\label{sec:isothermic}
\begin{figure}
\begin{center}
\includegraphics{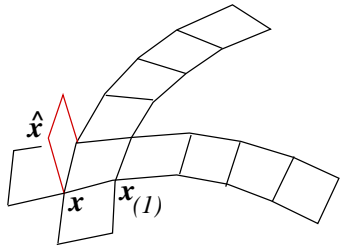}
\end{center}
\caption{Two intersecting initial strips of a two dimensional generalized
isothermic lattice allow to build all the lattice. The additional transverse
quadrilateral allows to build the lattice (together wih its Darboux transform)
in a three dimensional fashion} 
\label{fig:isoth-init}
\end{figure}

In the previous Sections we were mainly interested in generalized isothermic
lattices of dimension greater then two. However, simultaneous application of
the B-constraint and the quadratic constraint lowers dimensionality of the
lattice (in the sense of the initial boundary value problem). One can see it
from Figure~\ref{fig:twotothree}, which implies that two intersecting strips
made of planar quadrialterals with vertices in a quadric (see
Figure~\ref{fig:isoth-init}) can be extended to a
two dimensional quadrilateral lattice in the quadric. Because of
Lemma~\ref{lem:half-hex} such lattice satisfies Steiner's
version of the cross-ratio constraint \eqref{eq:isoth-cr}.  
One can define however geometrically two dimensional generalized isothermic
lattices (generalized discrete isothermic surfaces) without using the
three dimensional construction. An important tool here is the projective
interpretation of the discrete Moutard equation \cite{DGNS} as representing
quadrilateral lattice with additional linear relation between
any of its points $x$ and its four second-order neighbours $x_{(\pm 1 \pm 2)}$.
Geometrically, such five points of a
\emph{two dimensional B-quadrilateral lattice} 
are contained in a subspace of dimension three;
for generic two dimensional quadrilateral lattice such points are 
contained in a subspace of dimension four. To exclude further degenerations we
assume that no of the four points $x_{\pm 1}$, $x_{\pm 2}$ belongs to that three
dimensional subspace.
\begin{Def}
A two dimensional B-quadrilateral lattice in a hyperquadric 
$x:\ZZ^2\to\mathcal{Q}\subset\PP^M$
satisfying the local irreducibility condition is called a 
\emph{generalized discrete isothermic surface}.
\end{Def}
\begin{Rem}
Notice that the above Definition gives (the conclusion
was drawn by Alexander Bobenko) a geometric characterization of the
classical discrete isothermic surfaces of Bobenko and Pinkall. Mainly, 
the non-trivial intersection of a three dimensional subspace with 
the M\"{o}bius sphere
is a two dimensional sphere. After the stereographic projection,
which preserves co-sphericity of points, a discrete isothermic surface in
the M\"{o}bius (hyper)sphere gives \emph{circular two dimensional lattice
$s:\ZZ^2\to\EE^M$ such that for
any of its points $s$ there exists a sphere containig the point
and its four second-order neighbours $s_{(\pm 1 \pm 2)}$}.
\end{Rem}
Notice that, actually, all calculations where we used simultaneously both the
discrete Moutard equation and the quadratic constraint (algebraic proofs of
Proposition~\ref{prop:cross-ratio-gen}, Theorem~\ref{th:Darboux} and
Proposition~\ref{prop:superp}) remain true for $N=2$. Therefore the
corresponding results on the cross-ratio characterization of generalized
discrete isothermic surfaces, their Darboux transformation and the Bianchi
superposition principle are still valid. 

\begin{figure}
\begin{center}
\includegraphics{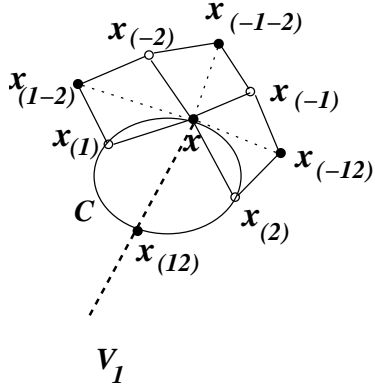}
\end{center}
\caption{Geometric construction of generalized discrete isothermic surfaces} 
\label{fig:geom-constr-surf}
\end{figure}

To complete this Section let us present the geometric construction of a
generalized discrete isothermic surface (see
Figure~\ref{fig:geom-constr-surf}). 
The basic step of the construction, which allows to build the generalized
discrete isothermic surface from two initial quadrilateral strips in a quadric
in a two dimensional fashion can be discribed as follows. Consider 
the four dimensional subspace 
$V_4 = \langle x, x_{(1)}, x_{(2)}, x_{(-1)}, x_{(-2)} \rangle$, where 
the basic step takes place. Denote by 
$V_3 = \langle x, x_{(-1-2)}, x_{(-12)}, x_{(1-2)} \rangle$ its three
dimensional subspace passing through the points $x$, $x_{(-1-2)}$, $x_{(-12)}$
and $x_{(1-2)}$, and by $V_2 = \langle x, x_{(1)}, x_{(2)} \rangle$ the plane of
the elementary quatrilateral whose fourth vertex $x_{12}$ we are going to find. 
In the
construction of the two dimensional B-quadrilateral lattice the vertex must
belong to the line $V_1 = V_3 \cap V_2$. In our case it should also belong to
the conic $\mathcal{C}=V_2 \cap \mathcal{Q}$. Because the conic contains 
already one
point $x$ of the line $V_1$, the second point  is unique. Notice that 
although the points
$x_{(-1)}$ and $x_{(-2)}$ do not play any role in the construction, they 
can be easily
recovered in a similar way as above.

\section{Conclusions and discussion}

In the paper we defined new integrable reduction of the lattice of planar
quadrilaterals, which contains as a particular example the discrete 
isothermic surfaces. We studied, by using geometric and algebraic means, various
aspects of such generalized isothermic lattices. In particular, we defined the
(analogs of the) Darboux transformations for the lattices and we showed the
corresponding permutablity principle.   

The theory of integrable systems is deeply connected with results of geometers
of the turn of XIX and XX centuries. The relation of integrability and geometry
is even more visible on the discrete level, where into the game there
enter basic results of the projective geometry. In our presentation of the
generalized isothermic lattices the basic geometric results were a variant of
the M\"{o}bius theorem and the generalization of the Miquel theorem to arbitrary
quadric, which combined together gave the corresponding generalization of the
Clifford theorem (known already to Miquel). 
An important tool in our research was also Steiner's description 
of conics and the geometric properties of von Staudt's algebra 
(see Appendix~\ref{sec:cr}).

\section*{Acknowledgements}
I would like to thank to Jaros{\l}aw Kosiorek and Andrzej Matra\'{s} for discusions
cencerning incidence geometry and related algebraic questions.
The main part of the paper was
prepared during my work at DFG Research Center MATHEON in 
Institut f\"{u}r Mathematik of the Technische Universit\"{a}t Berlin. 
The paper was supportet also in part by the Polish Ministry of  
Science and Higher Education research grant 1~P03B~017~28. 
Finally, it is my pleasure to thank the organizers of the SIDE VII Conference 
for support.

\appendix
\section{The cross-ratio and the projective structure of a conic}
\label{sec:cr}
For convenience of the Reader we have collected some facts from projective
geometry (see, for example \cite{Samuel,VeblenYoung}) used in the main text of
the paper.
Let $a,b,c$ be distinct points of the projective line $D$ over the field $\KK$. Given
$d\in D$, the
cross-ratio $\qcr(a,b;c,d)$ of the four points $a,b,c,d$ is defined as $h(d)\in
\hat\KK=\KK\cup\{\infty\}$ where $h$ is the unique projective transformation 
$D\to\hat\KK$ that takes $a$, $b$ and $c$ to $\infty$, $0$ and $1$, respectively.
For $D=\hat\KK$, with the usual conventions about operations with $0$ and $\infty$, the
cross-ratio is given by
\begin{equation}
\qcr(a,b;c,d) = \frac{(c-a)(d-b)}{(c-b)(d-a)}.
\end{equation}
Denote by $\boldsymbol{a}$ and $\boldsymbol{b}$ homogenous coordinates 
of the points
$a$ and $b$. If the homogenous coordinates of the points
$c$ and $d$ collinear with $a$, $b$ are, respectively, 
$\boldsymbol{c}=\alpha\boldsymbol{a}+ \beta\boldsymbol{b}$ and
$\boldsymbol{d}=\gamma\boldsymbol{a}+ \delta\boldsymbol{b}$, then 
\begin{equation} \label{eq:cr-proj}
\qcr(a,b;c,d) = \frac{\beta\gamma}{\alpha\delta}.
\end{equation}
Let $D$ and $D^\prime$ be projective lines, $a,b,c,d$ distinct points on $D$, and
$a^\prime,b^\prime,c^\prime,d^\prime$ distinct points on $D^\prime$. There exists a
projective transformation $u:D\to D^\prime$ taking $a,b,c,d$ into 
$a^\prime,b^\prime,c^\prime,d^\prime$, respectively, if and only if the cross-ratios
$\qcr(a,b;c,d)$ and  $\qcr(a^\prime,b^\prime;c^\prime,d^\prime)$ are equal.

\begin{figure}[!ht]
\begin{center}
\includegraphics{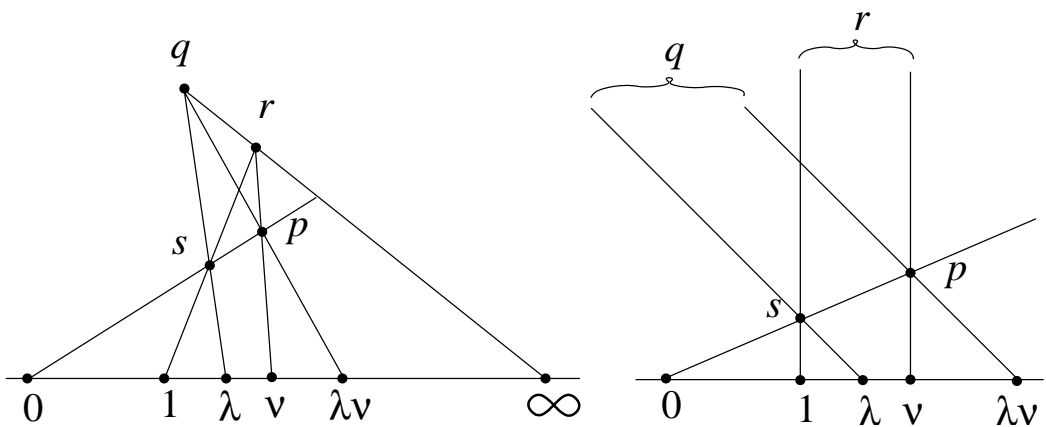}
\end{center}
\caption{Multiplication of cross-ratios on the line}
\label{fig:multiplication}
\end{figure}
Following von Staudt one can perform geometrically algebraic operations on
cross-ratios (see \cite{VeblenYoung} for details). We will be concerned with
geometric multiplication, which can be considered as "projectivization" of the
Thales theorem -- see the self-explanatory Figure~\ref{fig:multiplication}.

\begin{figure}
\begin{center}
\includegraphics{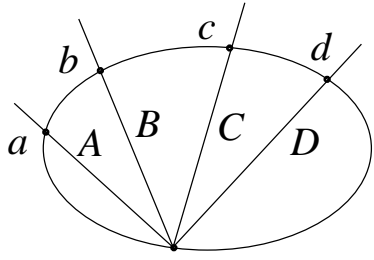}
\end{center}
\caption{The projective structure of a conic}
\label{fig:conic}
\end{figure}

The planar pencil of lines has the na\-tu\-ral projective structure 
inherited  from any line not intersecting its base.
Let $\mathcal{C}$ be an irreducible conic in a projective plane, and 
$a\in\mathcal{C}$
a point. To each line $D$ of the pencil $F_a$ of the base $a$, we associate 
the second
point where $D$ intersects $\mathcal{C}$
(see Figure~\ref{fig:conic}); we denote this point by $j_a(D)$. 
When $D$ is
the tangent to $\mathcal{C}$ at $a$, let $j_a(D)$ be the point $a$. Thus $j_a$ 
is a bijection from $F_a$ to $\mathcal{C}$.
If $b$ is another point on $\mathcal{C}$, the composition
$j_b^{-1}\circ j_a$ is a projective transformation from $F_a$ to $F_b$. 
Thus the
bijection from $F_a$ to $\mathcal{C}$ allows us to transport to $\mathcal{C}$ 
the
projective structure of $F_a$. This structure does not depend on the point~$a$.
Conversely, any projective transformation between two pencils defines a conic.

Finally we present the relation between the complex cross-ratio of the
M\"{o}bius geometry, and Steiner's conic cross-ratio.
\begin{Prop} \label{prop:Moebius-and-Steiner}
Four points $a,b,c,d\in\hat\CC$ are cocircular or collinear if and only if 
their
cross-ratio $\qcr(a,b;c,d)_\CC$ computed in 
$\hat\CC$, is real. The cross-ratio $\qcr(a,b;c,d)_\CC$ is 
equal to the cross-ratio $\qcr(a,b;c,d)$ computed using the real projective 
line structure
of the line or the circle considered as a conic.
\end{Prop}

\section{Auxiliary calculations}
\label{sec:aux-calc}
In this Appendix we would like to prove the "cross-ratio" characterization
(Proposition~\ref{prop:qred-cr}) of locally
irreducible quadrilateral lattices in quadrics, and then to give algebraic proof of the
basic Lemma~\ref{lem:half-hex}. It turns out that in the course of our 
calculations will give also
algebraic "down to earth" proofs of the
basic Lemmas~\ref{lem:gen-hex}, \ref{lem:BKP-hex} and
\ref{lem:qred-hex}.

Proposition~\ref{prop:qred-cr} is immediate consequence of the following result.
\begin{Lem} \label{lem:qred-hex-cr}
Under hypotheses of Lemma \ref{lem:qred-hex} and irreducibility
of the conics, being intersections of the planes of the quadrilaterals with the quadric,
the cross-ratios 
(defined with respect to the conics) 
of points on opposite sides of the hexahedron are connected by the following 
equation
\begin{equation} \label{eq:qred-cross-ratios} \begin{split}
\qcr(x_1,x_2;x_0,x_{12})& 
\, \qcr(x_{13},x_{23};x_3,x_{123}) 
\qcr(x_2,x_3; x_0,x_{23}) \, \qcr(x_{12},x_{13};x_1,x_{123}) = \\ 
&=\qcr(x_1,x_2;x_0,x_{13}) \, \qcr(x_{13},x_{23};x_2,x_{123}).
\end{split} 
\end{equation}
\end{Lem}
\begin{proof}
Let us choose the points $x_0$, $x_1$, $x_2$ and $x_3$ as the basis of
projective coordinate system in the corresponding three dimensional subspace, 
i.e.,                 
\begin{equation*}
\bx_0 = [1:0:0:0], \quad \bx_1 = [0:1:0:0],
\quad \bx_2 = [0:0:1:0],\quad \bx_3 = [0:0:0:1],
\end{equation*}
and denote by $[t_0 : t_1 : t_2 : t_3 ]$ the corresponding homogeneous
coordinates. 
For generic points $x_{ij}\in\langle x_0, x_{i}, x_{j} \rangle$, $1\leq i < j
\leq 3$, with homogeneous coordinates
\begin{equation*}
\bx_{12} =  [a_0:a_1:a_2:0], \quad
\bx_{13} =  [b_0:b_1:0:b_3], \quad
\bx_{23} =  [c_0:0:c_2:c_3],
\end{equation*}
one obtains, via the standard linear algebra, equations of
the planes $\langle x_1, x_{12}, x_{13} \rangle$, 
$\langle x_2, x_{12}, x_{23} \rangle$, $\langle x_3, x_{13}, x_{23} \rangle$
respectively,
\begin{align}
a_2 b_3 t_0 = & a_0 b_3 t_2 + b_0 a_2 t_3,\\
a_1 c_3 t_0 = & a_0 c_3 t_1 + c_0 a_1 t_3,\\
\label{eq:til_pi_12}
b_1 c_2 t_0 = & b_0 c_2 t_1 + c_0 b_1 t_2.
\end{align}
The intersection point $x_{123}$ of the planes has the following
coordinates $\bx_{123}=[y_0:y_1:y_2:y_3]$
\begin{equation} \label{eq:x_123-gen} \begin{split}
y_0 = & a_0 b_0 c_0 
\left( \frac{1}{a_2 b_1 c_3} + \frac{1}{a_1 b_3 c_2} \right), \\
y_1 = &  \frac{b_0 c_0}{b_3 c_2} +  \frac{a_0 c_0}{a_2 c_3} -
\frac{c_0^2}{c_2 c_3}  , \\
y_2 = &  \frac{a_0 b_0}{a_1 b_3} +  \frac{b_0 c_0}{b_1 c_3} -
\frac{b_0^2}{b_1 b_3}  , \\
y_3 = &  \frac{a_0 c_0}{a_1 c_2} +  \frac{a_0 b_0}{a_2 b_1} -
\frac{a_0^2}{a_1 a_2}  .
\end{split} \end{equation}
Up to now we have not used the additional quadratic restriction, and what we
have done was just the algebraic proof of Lemma~\ref{lem:gen-hex}.

Any quadric $\cQ$ passing through $x_0$,  $x_1$, $x_2$ and 
$x_3$ must have equation of the form
\begin{equation} \label{eq:quadric}
a_{01}t_0t_1 + a_{02}t_0t_2 + a_{03}t_0t_3 +
a_{12}t_1t_2 + a_{13}t_1t_3 + a_{23}t_2t_3 = 0.
\end{equation}
The homogeneous coordinates of the points $x_{12}$, $x_{23}$ and $x_{13}$ can be
parametrized in terms of the corresponding cross-ratios
$\lambda = \qcr(x_1,x_2;x_0,x_{12})$, $\nu = \qcr(x_2,x_3;x_0,x_{23})$ and 
$\mu = \qcr(x_1,x_3;x_0,x_{13})$ as
\begin{align} \label{eq:x12}
\bx_{12} &= 
\Big[\, \frac{\lambda a_{12}}{1-\lambda} \,:\, -\lambda a_{02} \,:\, 
a_{01}\,: \,0 \, \Big], \\
\label{eq:x23}
\bx_{23} &=  
\Big[\, \frac{\nu a_{23}}{1-\nu}\,: \, 0\,: \, -\nu  a_{03} \,: \,a_{02}
\,\Big],\\
\label{eq:x_13-quadr} 
\bx_{13} &= 
\Big[\, \frac{\mu a_{13}}{1-\mu} \,: \, -\mu  a_{03}\,: \, 0 \,: \, a_{01}
\,\Big].
\end{align}
We will only show how
to find the homogeneous coordinates of $x_{12}$ in terms of $\lambda$. 
Let us parametrize points of the conic
\begin{equation} \label{eq:conic}
a_{01}t_0t_1 + a_{02}t_0t_2 +  a_{12}t_1t_2 = 0,
\end{equation}
being intersection of the quadric \eqref{eq:quadric} with the plane $t_3=0$, 
by the planar pencil with base at $x_0$.
The point $x_0$ corresponds to the tangent to the conic at $x_0$ 
\begin{equation*}
a_{01}t_1 + a_{02}t_2 = 0,
\end{equation*}
while the points $x_1$ and $x_2$ correspond to lines $t_2=0$ and $t_1 = 0$,
respectively. The line 
$\langle x_0, x_{12} \rangle$ must have equation (see equation
\eqref{eq:cr-proj} in Appendix~\ref{sec:cr}) of the form
\begin{equation*}
a_{01}t_1 + \lambda a_{02}t_2 = 0,
\end{equation*}
which inserted into equation \eqref{eq:conic} of the conic gives, after exluding
the point $x_0$, the homogeneous
coordinates of the point $x_{12}$.

Inserting expressions \eqref{eq:x12}, \eqref{eq:x23} and \eqref{eq:x_13-quadr}
into formulas \eqref{eq:x_123-gen} we obtain homogeneous coordinates
$[y_0:y_1:y_2:y_3]$ of the point $x_{123}$ parametrized in terms of the
cross-ratios $\lambda$, $\nu$, $\mu$
\begin{equation} \label{eq:x_123-quadr} 
\begin{split}
y_0 = & \frac{a_{12}a_{23}a_{13}}{a_{01} a_{02} a_{03}} 
\frac{\lambda \nu - \mu}{(1-\lambda)(1-\mu)(1-\nu)}, \\
y_1 = &  \frac{a_{23}}{1-\nu} \left( 
\frac{a_{13}}{a_{01}a_{03}}\frac{\mu}{1-\mu} -
\frac{a_{23}}{a_{02}a_{03}}\frac{\nu}{1-\nu} -
\frac{a_{12}}{a_{01}a_{02}}\frac{\lambda\nu}{1-\lambda} \right)  , \\
y_2 = & \frac{a_{13}}{1-\mu} \left( 
-\frac{a_{13}}{a_{01}a_{03}}\frac{\mu}{1-\mu} +
\frac{a_{23}}{a_{02}a_{03}}\frac{\nu}{1-\nu} +
\frac{a_{12}}{a_{01}a_{02}}\frac{\mu}{1-\lambda} \right) , \\
y_3 = &  \frac{a_{12}}{1-\lambda} \left( 
\frac{a_{13}}{a_{01}a_{03}}\frac{\lambda}{1-\mu} -
\frac{a_{23}}{a_{02}a_{03}}\frac{1}{1-\nu} -
\frac{a_{12}}{a_{01}a_{02}}\frac{\lambda}{1-\lambda} \right) .
\end{split} 
\end{equation}
One can check that such expressions do satisfy the quadric equation 
\eqref{eq:quadric},
i. e., we have obtained the direct proof of Lemma~\ref{lem:qred-hex} under additional
assumption of ireducibility of the conics.

Further calculations give the cross-ratios on remaining three sides of the 
hexahedron 
\begin{align} \label{eq:qred-cross-ratios-opposite}
\qcr(x_{13}, x_{23}; x_3, x_{123}) & = 
 - \frac{\mu (1-\nu) a_{13} y_1}{\nu (1-\mu) a_{23} y_2} , \nonumber \\
\qcr(x_{12}, x_{13}; x_1, x_{123}) & =
- \frac{(1-\mu) a_{12} y_2}{(1-\lambda) a_{13} y_3}, \\ 
\qcr(x_{12}, x_{23}; x_2, x_{123}) & = 
\lambda \frac{(1-\nu) a_{12} y_1}{(1-\lambda) a_{23} y_3}, \nonumber
\end{align}
with $y_i$ given by \eqref{eq:x_123-quadr}, which implies 
equation \eqref{eq:qred-cross-ratios}.
\end{proof}
\begin{proof}[Algebraic proof of Lemma~\ref{lem:half-hex}]
We can express coplanarity of four points in $\PP^3$ as vanishing of the
determinant of the matrix formed by their homogeneous coordinates. In notation
of the proof above (up to equation \eqref{eq:x_123-gen}) we have
\begin{equation}
y_0 = \det \left(
\bx_{123} \: \bx_{1} \: \bx_{2} \: \bx_{3} \right) = 
 \frac{a_0b_0c_0}{a_1 a_2 b_1  b_3 c_2 c_3} 
\det \left( 
\bx_0 \: \bx_{12} \: \bx_{23} \: \bx_{13}  \right),
\end{equation}
which implies the statement of Lemma~\ref{lem:BKP-hex}.

Finally, when vertices of the hexahedron are contained in the quadric, condition $y_0=0$
inserted into equation \eqref{eq:x_123-quadr} gives $\mu = \lambda \nu$. 
\end{proof}

\bibliographystyle{amsplain}

\end{document}